\documentclass{PoS}
\usepackage{epsfig}
\usepackage{amsmath}
\usepackage{graphicx}
\usepackage{amssymb}
\def\beq{\begin{equation}}
\def\eeq{\end{equation}}
\def\bea{\begin{eqnarray}}
\def\eea{\end{eqnarray}}
\def\bwt{\begin{widetext}}
\def\ewt{\end{widetext}}
\def\nn{\nonumber}

\def\kbar{{\bar K}^0}
\def\bd{B^0}
\def\bdbar{{\bar B}^0}

\def\order{\lower 1.8ex \hbox{\LARGE\~{}}}

\def\btokpipi{B \to K \pi \pi}
\def\btokkk{B \to KK{\bar K}}
\def \babar{B{\sc a}B{\sc ar}}

\title{Measurement of $\gamma$ from three-body $B$ decays}

\ShortTitle{Measurement of $\gamma$ from three-body $B$ decays}

\author{Bhubanjyoti Bhattacharya and David London\\
				Physique des Particules, Universit\'e de Montr\'eal,\\ 
        C.P. 6128, succ.\ centre-ville, Montr\'eal, QC, Canada H3C 3J7\\
        E-mail: \email{bhujyo@lps.umontreal.ca}, \email{london@lps.umontreal.ca}}

\author{\speaker{Maxime Imbeault}\\
				D\'epartement de physique, C\'egep de Saint-Laurent,\\ 
        625, avenue Sainte-Croix, Montr\'eal, QC, Canada H4L 3X7\\
        E-mail: \email{mimbeault@cegep-st-laurent.qc.ca}}

\abstract{Using the \babar\ measurements of the Dalitz plots for the
  decays $\bd \to K^+\pi^0\pi^-$, $\bd \to K^0\pi^+\pi^-$, $B^+ \to
  K^+\pi^+\pi^-$, $\bd \to K^+ K^0 K^-$, and $\bd \to K^0 K^0 \kbar$,
  we demonstrate that it is possible to cleanly extract the weak phase
  $\gamma$.  An advantage of this Dalitz-plot method is that one can
  obtain many independent measurements of $\gamma$, thereby reducing
  its statistical error. An accurate determination of the errors,
  however, requires detailed knowledge of the data.  Hopefully, an
  experimental group will repeat the analysis, and obtain precise
  values of the errors.}

\FullConference{The European Physical Society Conference on High Energy Physics -EPS-HEP2013\\
		18-24 July 2013\\
		Stockholm, Sweden}

\begin{document}

\section{Introduction}
\label{sect:introduction}

In $B$ physics, the standard way of testing the standard model (SM) is
by measuring the three angles $\alpha$, $\beta$ and $\gamma$ of the
unitarity triangle in a number of different processes, and looking for
inconsistencies. Now, despite all the efforts that were done in this
vein, established direct measurements of $\gamma$ based on two-body
$B$ decays still suffer from large experimental uncertainties.  In
this talk, we present a method for implementing new constraints on
$\gamma$ using the three-body decays $\btokpipi$ and $\btokkk$.  The
content is based on Ref.~\cite{Bhattacharya:2013cla}.

For the general problem of extracting a weak phase from a decay $B \to
f$, the ideal candidate is a process for which (i) the final state $f$
is a CP eigenstate and (ii) the amplitude contains a single
contribution, so that there is no pollution due to interferences.  In
such a case, the measurement of the time-dependent CP asymmetry is
sufficient to cleanly extract the corresponding weak phase.  A perfect
example of this is the golden mode $\bd \to J/\psi K_s$, from which
the angle $\beta$ is obtained with good precision.  There is no such
prototype process for three-body $B$ decays, and a certain number of
general problems are encountered in extracting any weak-phase
information from these modes.  First, the final state is not, in
general, a CP eigenstate.  As an example, for the decay $\bd \to K_s
\pi^\pm \pi^\mp$, $f$ can be either even or odd under CP, depending on
the state of relative angular momentum of the pair of pions.  This is
a major complication because it means that one needs first to separate
the CP states of a given process before extracting any weak-phase
information from the data.  This separation of CP components can be
done by constructing an amplitude that is symmetric or antisymmetric
under the exchange of the two pions.  In section \ref{sect:method}, we
describe how this can be performed using the results of Dalitz-plot
fits.

A second problem is the fact that the amplitude of a three-body $B$
decay process usually involves several different contributions.  It
was shown in Ref.~\cite{Lorier:2010xf} that any such decay can be
expressed in terms of ten graphical diagrams (neglecting
power-suppressed contributions from weak-annihilation topologies):
gluonic penguins (${\tilde P}'_{uc}$ and ${\tilde P}'_{tc}$),
color-allowed trees ($T'_{1,2}$), color-suppressed trees ($C'_{1,2}$),
color-allowed electroweak-penguins (EWP's) ($P'_{EW1,2}$) and
color-suppressed EWP's ($P_{EW1,2}^{\prime C}$).  This large number of
diagrams implies that any extraction of weak-phase information from
three-body $B$ decays has to deal with a large number of hadronic
theoretical parameters.  Thus, a single mode is not
sufficient. Instead, one must consider a set of several decays, all
related by a flavor symmetry and written in terms of the same
diagrams. In the current method, assuming flavor-SU(3) symmetry, the
following five decays are considered: $\bd \to K^+\pi^0\pi^-$, $\bd
\to K^0\pi^+\pi^-$, $\bd \to K^+ K^0 K^-$, $\bd \to K^0 K^0 \kbar$ and
$B^+ \to K^+\pi^+\pi^-$.  The last one is not required for extracting
$\gamma$ under full SU(3), but we use it in order to probe the
leading-order effect of SU(3) breaking.  This is described in
detail in the following sections.

There are many unknown hadronic parameters to deal with, but it was
shown in Ref.~\cite{Imbeault:2010xg} that, within flavor-SU(3)
symmetry, tree and EWP diagrams are not independent, to a good
approximation. In the current method we use this fact to reduce the
number of hadronic parameters.  However, the tree-EWP relations
require the final state to be fully symmetric.  In order to use these
relations, and because of the CP-eigenstate problem discussed above,
all final states considered here have to be fully symmetrized (and not
symmetrized only with respect to the pair of pions or kaons).

Third, contrary to two-body decays, the amplitude of a three-body
process is momentum dependent.  Thus, all diagrams are explicit
functions of the momenta of the final-state particles, so that one cannot get values of gamma and all the hadronic parameters from a global fit to integrated observables.  The thing that has to be done is a fit for a given bin in
the momentum space and to repeat it for all possible bins.  Several
values of $\gamma$ are then obtained and need to be averaged
appropriately.

All of these problems can be overcome but they still imply technical
difficulties in the implementation of the method, thus leading to
numerical uncertainties in our final results.  These cannot be
estimated rigorously without full access to the data, so that all
numerical results presented in this talk should be regarded more as a
pedagogical example than a firm state-of-the-art extraction of
$\gamma$.  Our aim is to demonstrate the principle of the method with
the hope that experimentalists can apply it directly to the data.

\section{Description of the method}
\label{sect:method}

The procedure for extracting $\gamma$ from $\btokpipi$ and $\btokkk$
decays can be summarized in four steps\footnote{In the talk they
  were presented as five steps, but it is more simple and compact in
  the written version to merge two of them.}.  The first step is to
construct a fully-symmetric amplitude from the Dalitz-plot
results. Consider a general decay mode $B \to P_1 P_2 P_3$.  One
defines the three Mandelstam variables $s_{ij} \equiv (p_i+p_j)^2$,
where $p_i$ is the four-momentum of $P_i$.  Only two of them are
independent since they obey the kinematic relation
$s_{12}+s_{23}+s_{13} = m_B^2 + m_1^2 + m_2^2 + m_3^2$, where $m_B$ is
the mass of the $B$ meson and $m_i$ is the mass of $P_i$.  The
amplitude of the three-body decay is therefore basically a complex
function of two Mandelstam variables: $A(s_{12}, s_{13})$. In
experimental Dalitz-plot analyses, the isobar model is usually assumed
for $A(s_{12}, s_{13})$ in a fit of the event distribution. Within
this framework, the amplitude is written as a sum of complex terms
describing each of the resonant and nonresonant components of the
decay.  Explicitly,
\beq
A(s_{12}, s_{13}) = \mathcal{N} \sum_{j} c_j e^{i \theta_j} F_j (s_{12},s_{13})~.
\label{eq:isobar}
\eeq
Above, the $c_j e^{i \theta_j}$ are complex coefficients obtained from
the experimental fit, the $F_j$ are the dynamical functions of the
model, and $\mathcal{N}$ is a global normalization constant fixed by
the requirement that the integrated Dalitz plot reproduces the correct
experimental branching fraction of the decay.  The important point is
that, within these assumptions, the amplitude and its momentum
dependence are fully measured by the fit.  (On the other hand, the use
of the isobar coefficients implies a model dependence of the output.)
It is now straightforward to construct the fully-symmetric amplitude
$A_{\rm fs}(s_{12},s_{13})$ that has the correct CP properties
\cite{ReyLeLorier:2011ww}.  Explicitly, it takes the form
\bea
A_{\rm fs}(s_{12},s_{13})&=& \frac{1}{\sqrt{6}} \left( A(s_{12},s_{13}) + A(s_{13},s_{12}) + A(s_{12},s_{23}) \right.\nn\\
&&\left. +~A(s_{23},s_{12}) + A(s_{13},s_{23}) + A(s_{23},s_{13}) \right)~.
\eea

The second step is to construct a set of fully-symmetric observables.
In analogy with the standard branching fraction, direct CP asymmetry
and indirect CP asymmetry, one can define the following fully-symmetric
observables for any given value of momenta of the final-state
particles:
\bea
X(s_{12},s_{13}) &=& |A_{\rm fs}(s_{12},s_{13})|^2 + |{\bar A}_{\rm fs}(s_{12},s_{13})|^2\nn\\
Y(s_{12},s_{13}) &=& |A_{\rm fs}(s_{12},s_{13})|^2 - |{\bar A}_{\rm fs}(s_{12},s_{13})|^2\nn\\
Z(s_{12},s_{13}) &=& {\rm Im}\left( A^*_{\rm fs}(s_{12},s_{13}) {\bar A}_{\rm fs}(s_{12},s_{13}) \right) ~,
\label{eq:observables}
\eea
where the observable $Z(s_{12},s_{13})$ applies only to those decays
for which the final state is accessible to both the $\bd$ and the
$\bdbar$ mesons. For the five decays we consider in the method we then
have, in principle, 13 fully-symmetric observables: five $X$'s, five
$Y's$ and three $Z's$ (for $\bd \to K^0\pi^+\pi^-$, $\bd \to K^+ K^0
K^-$ and $\bd \to K^0 K^0 \kbar$).  This is the set of observables we
use to fit for $\gamma$.

In step three, we obtain a theoretical expression for each of the
observables.  This is done by writing the fully-symmetric amplitude of
each of the five considered decays using graphical diagrams as
described in Ref.~\cite{Lorier:2010xf}.  After using the tree-EWP
relations and combining diagrams that always appear in the same
linear combination, we get the following expressions:
\bea
\label{eq:effamps}
2 A(\bd \to K^+\pi^0\pi^-)_{\rm fs} &=& b e^{i\gamma} - \kappa c ~, \nn\\
\sqrt{2} A(\bd \to K^0\pi^+\pi^-)_{\rm fs} &=& -d e^{i\gamma} - {\tilde P}'_{uc} e^{i\gamma} - a + \kappa d ~, \nn\\
\sqrt{2} A(B^+ \to K^+ \pi^+ \pi^-)_{\rm fs} &=& -c e^{i\gamma} -{\tilde P}'_{uc} e^{i\gamma} - a + \kappa b ~, \nn\\
\sqrt{2} A(\bd \to K^+ K^0 K^-)_{\rm fs} &=& \alpha_{SU(3)} (-c e^{i\gamma} -{\tilde P}'_{uc} e^{i\gamma} - a + \kappa b ) ~, \nn\\
A(\bd \to K^0 K^0 \kbar)_{\rm fs} &=& \alpha_{SU(3)} ({\tilde P}'_{uc} e^{i\gamma} + a ) ~,
\eea
where $a$, $b$, $c$ and $d$ are four effective diagrams defined as
\beq
a \equiv - {\tilde P}'_{tc} + \kappa \left(\frac23 T'_1 + \frac13 C'_1
+ \frac13 C'_2 \right) ~,~~
b \equiv T'_1 + C'_2 ~,~~
c \equiv T'_2 + C'_1 ~,~~
d \equiv T'_1 + C'_1 ~,
\label{eq:effdiag}
\eeq
%
and $\alpha_{SU(3)}$ is a complex number parametrizing the
leading-order SU(3) breaking between $\btokpipi$ and $\btokkk$ decays.

There are several important comments regarding the above two
equations.  First, even if the momentum dependence is not written
explicitly, all of the above diagrams are functions of the momenta of
the final-state particles.  Second, $A(B^+ \to K^+ \pi^+ \pi^-)_{\rm
  fs}$ and $A(\bd \to K^+ K^0 K^-)_{\rm fs}$ are identical
under flavor-SU(3) symmetry.  This can be seen explicitly in
Eq.~(\ref{eq:effamps}) by setting $\alpha_{SU(3)} = 1$.  Although
$\alpha_{SU(3)}$ is a complex number, its phase always cancels when
calculating fully-symmetric observables as in
Eq. (\ref{eq:observables}), so that its inclusion effectively adds
only a single real parameter in the counting.  Finally, it should be
mentioned that approximate SU(3) relations were used in order to write
all EWP's in terms of trees in Eqs.~(\ref{eq:effamps}) and
(\ref{eq:effdiag}).  These relations imply a direct proportionality
$P'_{EWi} = \kappa T'_i$, where
\bea
\kappa \equiv - \frac{3}{2} \frac{|\lambda_t^{(s)}|}{|\lambda_u^{(s)}|}
\frac{c_9+c_{10}}{c_1+c_2} ~,
\eea
and where $c_i$'s are Wilson coefficients of the effective
electroweak hamiltonian and $\lambda_p^{(s)}=V^*_{pb} V_{ps}$.

{}From the above, for each point in the momentum space there are, in
principle, 11 theoretical parameters: 9 hadronic parameters (five
magnitudes of ${\tilde P}'_{uc}$, $a$, $b$, $c$, $d$ and their four
relative strong phases), $|\alpha_{SU(3)}|$ and $\gamma$.  Thus, in
step four, $\gamma$ can be extracted for any given bin of the momentum
space and all these results can be averaged.  In the next section we
give a numerical example of how this can be done in practice.

\section{Numerical example}
\label{sect:numerical}

In order to demonstrate how this method can be applied concretely to
data, we use the published results of Dalitz-plot analyses from the
\babar\ collaboration \cite{expt1,expt2,expt3,expt4,expt5}.  In these
papers, event distributions for the five decays of interest were
fitted within the isobar model, and the complex coefficients $c_j e^{i
  \theta_j}$ of Eq.~(\ref{eq:isobar}) extracted.  As described in the
previous section, we can then construct fully-symmetric amplitudes and
a set of fully-symmetric observables for any given point in the
momentum space.  But it is important to note that in four of these
five papers, the isobar coefficients are quoted with statistical
errors only. Therefore the following results \textit{do not} include
systematic effects, which are not negligible in general.  In order to
estimate the error bars of calculated fully-symmetric observables, we
let all input isobar parameters vary within their 1$\sigma$ ranges and
include correlations among these input numbers whenever this
information is provided.  There is another important remark about the
set of input numbers.  In Ref.~\cite{expt5}, $A(\bd \to K^0 K^0
\kbar)$ was assumed to be equal to its CP-conjugate amplitude. In
order to be consistent with this we have set the gluonic penguin
diagram ${\tilde P}'_{uc}$ to zero, which reduces the number of
theoretical parameters by two.  But, consistently, the CP-violating
observables $Y$ and $Z$ in this mode also vanish.  This reduces the
number of observables by two so that there is no change in the
counting balance.  Note that this approximation is not expected to
have a big impact on the output numbers since this penguin diagram is
CKM suppressed and is expected to be sub-dominant compared to ${\tilde
  P}'_{tc}$.

\begin{figure}[!htbp]
\begin{center}
\includegraphics[width=0.4\textwidth]{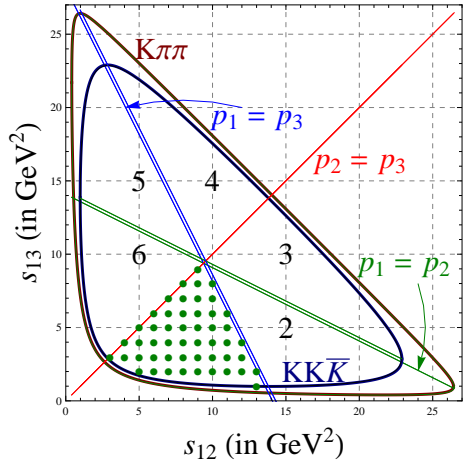}
\caption{Kinematic boundaries and symmetry axes of $\btokpipi$ and
$\btokkk$ Dalitz plots. The symmetry axes divide each plot into six
zones, five of which are marked 2-6. The fifty dots in the region
of overlap of the first of six zones from all Dalitz plots are used
for the $\gamma$ measurement.
\label{DPbdry}}
\end{center}
\end{figure}

Since we do not work directly with the event distributions, in this
numerical demonstration we do not perform a binning of the momentum
space.  Instead we consider a regular grid of 50 points with a
resolution of 1 GeV$^2$.  This is represented in Fig.~\ref{DPbdry}.
Note also that, since the observables are fully symmetric by
construction, the sample points are restricted to one sixth of the
Dalitz plot in order to avoid multiple counting.
At this point it is important to mention that, in the current
numerical analysis, we do \textit{not} include correlations among
observables at different points of the Dalitz plot.  They can, in
principle, be strongly correlated since they are calculated from the
same constructed amplitudes, and not derived directly from the data.
These correlations may have an important effect on the errors of the
final result.  Note also that the choice of resolution of 1 GeV$^2$ is
completely \textit{ad hoc}.  There is no formal justification for this
and it can also have a big effect on the final error since the error
of the average of the $N$ values of $\gamma$ extracted from $N$ points
of the Dalitz plot naively scales as $1/\sqrt{N}$.  A more rigorous
analysis needs to be performed directly from the data in order to
avoid these assumptions.

\begin{figure*}[!htbp]
\begin{center}
\includegraphics[width=0.8\textwidth]{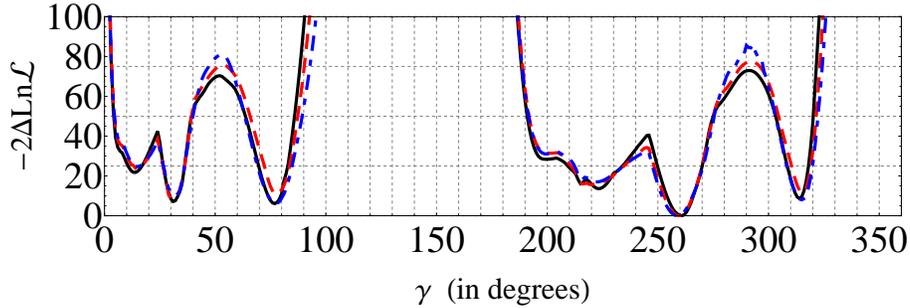} 
\caption{Results of the maximum-likelihood fits. The solid (black)
  curve represents the fit assuming flavor-SU(3) symmetry. The short
  dashes (red) represent the fit where flavor-SU(3) breaking is fixed
  by a point-by-point comparison of Dalitz plots for $B^+\to
  K^+\pi^+\pi^-$ and $B^0\to K^+ K^0 K^-$. The long dashes (blue)
  represent the fit with inputs from five Dalitz plots and an extra
  hadronic fit parameter $|\alpha_{SU(3)}|$. \label{maxlik}}
\end{center}
\end{figure*}

For each of the 50 points we perform a maximum-likelihood fit within
three scenarios (in order to test flavor-SU(3) breaking), and then
combine all likelihood distributions.  In the first scenario (Fit 1),
the mode $B^+ \to K^+ \pi^+ \pi^-$ is excluded and full SU(3) is
assumed by setting $|\alpha_{SU(3)}|=1$.  In the second scenario (Fit
2), $|\alpha_{SU(3)}|$ is calculated using the ratio of $X$'s from the
$\bd \to K^+ K^0 K^-$ and $B^+ \to K^+ \pi^+ \pi^-$ modes, and is then
used in the fit with the other modes.  Finally, in the last scenario
(Fit 3), $|\alpha_{SU(3)}|$ is left as a free parameter and all five
decays are fit simultaneously.  The combined $-2 \Delta \ln {\rm
  L}(\gamma)$ distributions for these three analyses are presented in
Fig. \ref{maxlik}.
All three scenarios produce very similar results, suggesting a small
SU(3) breaking.  There are four favored solutions, and this discrete
ambiguity cannot be resolved without additional outside input.  As an
example, the numerical results of $\gamma$ for Fit 1 are
$(31^{+2}_{-1})^\circ$, $(77 \pm 2)^\circ$, $(261^{+2}_{-3})^\circ$
and $(314 \pm 2)^\circ$.  It is interesting to note that only one
solution ($77^\circ$) is consistent with established direct
measurements of $\gamma$.  Let us emphasize again that the small error
bars obtained are a consequence of the choice of the 1 GeV$^2$ grid,
the fact that systematic effects are not included in the input
numbers, and the fact that we do not include correlations among
observables at different points of the Dalitz plots.  These numbers
are pedagogical examples containing several assumptions, but they are
still values of $\gamma$ extracted from data of three-body $B$ decays.
This proves that such an extraction \textit{can} be done. This is the
main point of this talk.

Finally, it is interesting to mention that the SU(3)-breaking
parameter $|\alpha_{SU(3)}|$ can be extracted for each point of the
Dalitz plot from the $A$'s and $\bar A$'s independently. Averaging
over the 50 points, we find $|\alpha_{SU(3)}| = 0.97 \pm 0.04$ ($A$'s)
or $0.99 \pm 0.04$ ($\bar A$'s). These are consistent with one another
and are also consistent with unity, again pointing towards a small
SU(3)-breaking effect. The bottom line is that, even if SU(3) breaking
is not under full control with the addition of a single breaking
parameter, all clues suggest a small effect.

\section{Conclusion}
\label{sect:conclusion}

In this talk, we have described a new method for extracting the angle
$\gamma$ from the data of $\btokpipi$ and $\btokkk$ decays, and we
have demonstrated numerically that such an extraction can be done.
All numbers presented here contain several simplifying assumptions, so
that they should be regarded mostly as a pedagogical example.  At
present, it is impossible to predict what would be the output of a
reproduction of this analysis performed directly from the data and
including all statistical and correlation effects.  But the fact is
that three-body $B$ decays can provide additional constraints on
$\gamma$.

\acknowledgments 
A special thank you goes to E. Ben-Haim for his input to this project.
This work was financially supported by NSERC of Canada (BB, DL) and by
FQRNT of Qu\'ebec (MI).  MI would also like to thank A. Soni for
helpful discussions during the conference and the organizers of the
EPS-HEP2013.

\end{document}